\documentclass[prb,aps,twocolumn,superscriptaddress]{revtex4-1}

\usepackage{epsfig}
\usepackage{amsmath}
\usepackage{amsthm}
\usepackage{amssymb,braket}
\usepackage{verbatim}
\usepackage{bm}
\usepackage{hyperref}
\usepackage[all]{hypcap}
\usepackage{here}
\usepackage{sidecap}
\usepackage{float}
\usepackage{graphicx}
\usepackage{epstopdf}

\newcommand{\bpm}{\begin{pmatrix}}
\newcommand{\epm}{\end{pmatrix}}
\newcommand{\ba}{\begin{eqnarray}}
\newcommand{\ea}{\end{eqnarray}}

\begin{document}
\title{Electronic band structure of (111) SrRuO$_{3}$ thin film$-$an  angle-resolved photoemission spectroscopy study}

\author{Hanyoung Ryu}
\affiliation{Center for Correlated Electron Systems, Institute for Basic Science (IBS), Seoul 08826, Republic of Korea}
\affiliation{Department of Physics and Astronomy, Seoul National University (SNU), Seoul 08826, Republic of Korea}

\author{Yukiaki Ishida}
\affiliation{Center for Correlated Electron Systems, Institute for Basic Science (IBS), Seoul 08826, Republic of Korea}
\affiliation{Institute of Solid State Physics, The University of Tokyo, Kashiwa, Chiba 277-8581, Japan}

\author{Bongju Kim}
\affiliation{Center for Correlated Electron Systems, Institute for Basic Science (IBS), Seoul 08826, Republic of Korea}
\affiliation{Department of Physics and Astronomy, Seoul National University (SNU), Seoul 08826, Republic of Korea}

\author{Jeong Rae Kim}
\affiliation{Center for Correlated Electron Systems, Institute for Basic Science (IBS), Seoul 08826, Republic of Korea}
\affiliation{Department of Physics and Astronomy, Seoul National University (SNU), Seoul 08826, Republic of Korea}

\author{Woo Jin Kim}
\affiliation{Center for Correlated Electron Systems, Institute for Basic Science (IBS), Seoul 08826, Republic of Korea}
\affiliation{Department of Physics and Astronomy, Seoul National University (SNU), Seoul 08826, Republic of Korea}

\author{Yoshimitsu Kohama}
\affiliation{Institute of Solid State Physics, The University of Tokyo, Kashiwa, Chiba 277-8581, Japan}

\author{Shusaku Imajo}
\affiliation{Institute of Solid State Physics, The University of Tokyo, Kashiwa, Chiba 277-8581, Japan}

\author{Zhuo Yang}
\affiliation{Institute of Solid State Physics, The University of Tokyo, Kashiwa, Chiba 277-8581, Japan}

\author{Wonshik Kyung}
\affiliation{Center for Correlated Electron Systems, Institute for Basic Science (IBS), Seoul 08826, Republic of Korea}
\affiliation{Department of Physics and Astronomy, Seoul National University (SNU), Seoul 08826, Republic of Korea}

\author{Sungsoo Hahn}
\affiliation{Center for Correlated Electron Systems, Institute for Basic Science (IBS), Seoul 08826, Republic of Korea}
\affiliation{Department of Physics and Astronomy, Seoul National University (SNU), Seoul 08826, Republic of Korea}

\author{Byungmin Sohn}
\affiliation{Center for Correlated Electron Systems, Institute for Basic Science (IBS), Seoul 08826, Republic of Korea}
\affiliation{Department of Physics and Astronomy, Seoul National University (SNU), Seoul 08826, Republic of Korea}

\author{Inkyung Song}
\affiliation{Center for Correlated Electron Systems, Institute for Basic Science (IBS), Seoul 08826, Republic of Korea}

\author{Minsoo Kim}
\affiliation{Center for Correlated Electron Systems, Institute for Basic Science (IBS), Seoul 08826, Republic of Korea}
\affiliation{Department of Physics and Astronomy, Seoul National University (SNU), Seoul 08826, Republic of Korea}

\author{Soonsang Huh}
\affiliation{Center for Correlated Electron Systems, Institute for Basic Science (IBS), Seoul 08826, Republic of Korea}
\affiliation{Department of Physics and Astronomy, Seoul National University (SNU), Seoul 08826, Republic of Korea}

\author{Jongkeun Jung}
\affiliation{Center for Correlated Electron Systems, Institute for Basic Science (IBS), Seoul 08826, Republic of Korea}
\affiliation{Department of Physics and Astronomy, Seoul National University (SNU), Seoul 08826, Republic of Korea}

\author{Donghan Kim}
\affiliation{Center for Correlated Electron Systems, Institute for Basic Science (IBS), Seoul 08826, Republic of Korea}
\affiliation{Department of Physics and Astronomy, Seoul National University (SNU), Seoul 08826, Republic of Korea}

\author{Tae Won Noh}
\email{twnoh@snu.ac.kr}
\affiliation{Center for Correlated Electron Systems, Institute for Basic Science (IBS), Seoul 08826, Republic of Korea}
\affiliation{Department of Physics and Astronomy, Seoul National University (SNU), Seoul 08826, Republic of Korea}

\author{Saikat Das}
\email {DAS.Saikat@nims.go.jp} 
\thanks{Present address: Research Center for Magnetic and Spintronic Materials, National Institute for Materials Science, 1-2-1 Sengen, Tsukuba 305-0047, Japan}
\affiliation{Center for Correlated Electron Systems, Institute for Basic Science (IBS), Seoul 08826, Republic of Korea}
\affiliation{Department of Physics and Astronomy, Seoul National University (SNU), Seoul 08826, Republic of Korea}

\author{Changyoung Kim}
\email{changyoung@snu.ac.kr}
\affiliation{Center for Correlated Electron Systems, Institute for Basic Science (IBS), Seoul 08826, Republic of Korea}
\affiliation{Department of Physics and Astronomy, Seoul National University (SNU), Seoul 08826, Republic of Korea}

\begin{abstract}
We studied the electronic band structure of pulsed laser deposition (PLD) grown (111)-oriented SrRuO$_3$ (SRO) thin films using \textit{in situ} angle-resolved photoemission spectroscopy (ARPES) technique. We observed previously unreported, light bands with a renormalized quasiparticle effective mass of about 0.8$m_{e}$. The electron-phonon coupling underlying this mass renormalization yields a characteristic ``kink" in the band dispersion. The self-energy analysis using the Einstein model suggests five optical phonon modes covering an energy range  44 to 90 meV contribute to the coupling. Besides, we show that the quasiparticle spectral intensity at the Fermi level is considerably suppressed, and two prominent peaks appear in the valance band spectrum at binding energies of 0.8 eV and 1.4 eV, respectively. We discuss the possible implications of these observations. Overall, our work demonstrates that high-quality thin films of oxides with large spin-orbit coupling can be grown along the polar (111) orientation by the PLD technique, enabling \textit{in situ} electronic band structure study. This could allow for characterizing the thickness-dependent evolution of band structure of (111) heterostructures$-$a prerequisite for exploring possible topological quantum states in the bilayer limit.

\end{abstract}
\maketitle

Perovskite transition metal oxides (TMOs) encompass a wide variety of properties like high-temperature superconductivity, magnetism, ferroelectricity, metal-insulator transition, colossal magnetoresistance, and multiferroicity \cite{MIT_RMP, Multiferroic}. The plethora of physical properties in these materials originate from the subtle interplay among the charge, lattice, spin, and orbital degrees of freedom. Tweaking this interplay via epitaxy or heterointerfacing, furthermore, allows manipulating these properties and even designing novel phenomena or functionalities, which are unattainable by the bulk solid-state synthesis route. Examples include strain-induced enhancement of ferroelectricity and superconductivity \cite{BTO_Strain, LSCO_Strain}, high-mobility conducting interface \cite{2DEG_Eom, Harold_LAO/STO}, interface ferromagnetism, polar skyrmions \cite{Electrode_3, Skyr_Nano_Lett}. While the majority of these works have been carried out using heterostructures that are grown along the crystallographic [001] direction, their (111)-oriented counterparts are gaining considerable attention recently \cite{Okamoto_111, Okamoto_111_2, Okamoto_111_3, Mott}.

Perhaps the biggest motivation to study (111)-oriented TMO heterostructures stem from the prediction of stabilizing novel topological phases in the bilayer limit  \cite{Okamoto_111_2,Okamoto_111_3}. Specific to this orientation, the trigonal crystal field symmetry, together with a sizable spin-orbit coupling, is argued to open topologically protected energy gaps in an otherwise topologically trivial band structure. The strong electronic correlation that is inherent to the TMOs is further expected to enrich their topological properties. An essential step in this direction is first to comprehensively understand the band structure of thicker (111) TMO films, and subsequent characterization with thickness scaling. Thus, \textit {in situ} angle-resolved photoemission spectroscopy (ARPES) studies could be highly beneficial, which, however, requires overcoming difficulties involved growing high-quality thin films on the polar (111) surfaces. Besides, the requirement of strong spin-orbit coupling (SOC) strength further narrows the choice of materials to the TMOs that contain heavier elements. Accordingly, to our best knowledge, ARPES studies on (111) thin films are limited to the 3\textit{d} Nickelates \cite{FYBruno,Arab}, where the SOC strength is expected to be weak. It is, therefore, instructive also exploring TMOs with larger SOC.   

In this regard, SrRuO$_3$ (SRO)$-$a 4\textit{d} TMO is of particular interest since both the SOC strength (0.1-0.15 eV) and electronic correlation  are rather sizeable \cite{Mattheiss, DOS_Calculation}. In the bulk, SRO is an itinerant ferromagnet (below 165 K) and exhibits a Fermi-liquid behavior below 40 K \cite{Shen}. Thin films of SRO that are grown along the [001] direction have been extensively studied as a model system in the context of anomalous Hall effect originating from the magnetic monopole in the momentum space \cite{SRO_RMP, Monopole}. Recently, it has gained renewed interest due to the observation of the topological Hall effect \cite{Electrode_3}. Besides, they are commonly used as metallic electrodes-thanks to the feasibility of growing atomically smooth films with high crystalline quality. The electronic band structure of (001) SRO films is relatively well understood both on the theoretical \cite{DOS_Calculation, DFT_DMFT_Calculation, LDA_DMFT, SRO_Japan} and experimental fonts \cite{Origin_of_kink, Off_cation, Shen, SRO_CRO_ARPES}.

In contrast, the (111)-oriented SRO films have received moderate attention. Notably, (111) SRO thin films have been shown to exhibit anomalously enhanced magnetism (compared to the bulk), and conductivity compared to (001) SRO thin films \cite{Magnetism_SRO, RR_distance, Anistropy}. It is also proposed that (111) SRO heterostructures could support half-metallic ground state at room temperature and upon electron doping, a quantum anomalous hall state could arise in the bilayer limit \cite{QAH_SRO_111}. Both of these properties are highly relevant for spintronic applications. Despite these intriguing electromagnetic properties, the electronic band structure of (111) SRO films has not been studied experimentally \cite{QAH_SRO_111}.

In this rapid communication, we study the electronic band structure of (111) SRO thin film by means of \textit{in situ} APRES technique. We find the existence of a light band with a characteristic  renormalized quasiparticle effective mass of 0.8$m_{e}$. Based on the Einstein modeling of self-energy, we show that this mass renormalization can be attributed to the interaction between electrons and multiple phonon modes. Both the renormalized effective mass value, and the nature of the electron-phonon mode coupling differs from previously reported (001) SRO films, thereby highlighting the unique electronic property of (111) SRO film.

\begin{figure}[t]
\centering
\includegraphics[scale=0.75]{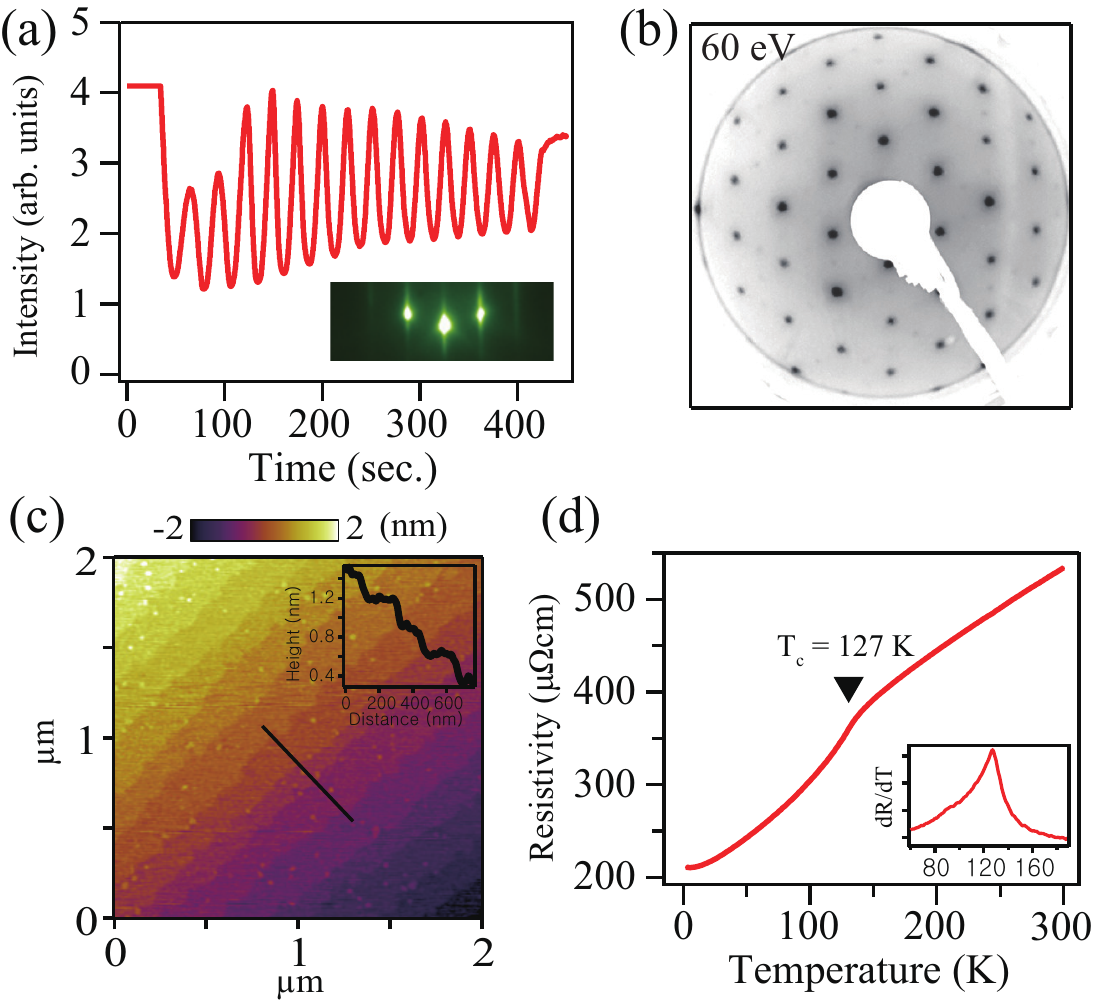}
\caption{(Color online) (a) RHEED image taken along the [1$\bar{1}$0] azimuth. (b) LEED image of 15 u.c$.$ SRO film, taken at an electron kinetic energy of 60 eV. (c) AFM image showing the step-terrace structure, and inset shows the height profile along the black solid line. (d) Temperature dependent resistivity of 15 u.c$.$ (111) SRO, and the inset shows the derivative plot highlighting the onset of ferromagnetism around 127 K.}
\label{fig:digraph}
\end{figure}


SRO thin films were grown on the B-site terminated (111) oriented SrTiO$_3$ (STO) substrates \cite{STO_growth}  using the pulsed laser deposition technique (KrF laser, $\lambda$ = 248 nm). During the growth, the substrate temperature and the background oxygen partial pressure were set to 680 $^{\circ}$C and 100 mTorr, respectively. Meanwhile, the laser fluence and repetition rate were fixed to 1.1 J/cm$^{2}$ and 1 Hz, respectively. The growth dynamics were monitored by the reflection high energy electron diffraction (RHEED) technique. After growth, samples were cooled down to room temperature, and the oxygen flow was stopped for achieving a high vacuum, $\sim$5$\times$10$^{-9}$ Torr. After growth, the samples were transferred to the preparation chamber and post-annealed in 1$\times$10$^{-9}$ Torr oxygen partial pressure at 510 $^{\circ}$C for 30 minutes to achieve a clean surface. Subsequently, the films were transferred \textit{in situ} to the angle-resolved photoemission spectroscopy (ARPES) chamber, which is equipped with VG Scienta DA30 analyzer and ultraviolet light source and monochromator from Fermi instrument. During the ARPES measurement, the base pressure in the chamber was better than 8$\times$10$^{-11}$ Torr, and the sample temperature was 10 K. For the ARPES measurement, we employed HeI (21.22 eV) light. X-ray photoemission spectroscopy (XPS) measurement was performed using Al K$\alpha$ photon (1486.6 eV) at room temperature in the XPS analyzer chamber equipped with SPECS XR50 X-ray photon source (Fig$.$ S1 in Supplemental Material \cite{supp}). After the ARPES and XPS measurements, the samples were characterized by low energy electron diffraction (LEED). Electrical transport measurement was performed by ultrasonically bonding gold wires on to the film in four-terminal configuration and using a Quantum Design PPMS. The surface morphology was probed using an Asylum Cypher atomic force probe microscope (AFM).


Figure 1(a) displays the characteristic RHEED intensity profile of the specular [00] Bragg reflex (inset of Fig$.$ 1(a)) during SRO thin film growth. The specular RHEED intensity exhibits clear oscillations, reflecting the layer-by-layer growth of SRO film. The RHEED intensity oscillations enable us to preciously controlling the film thickness, which we varied between 7-30 unit cells (u$.$c$.$). As a representative figure, here, we have shown the RHEED intensity profile and pattern taken during and after the growth of a 15 u$.$c$.$ thick SRO film. As shown in the inset of Fig$.$ 1(a), the RHEED pattern of the SRO film consists of sharp diffraction spots forming a Laue circle, which suggests coherent growth of crystalline domains with long-range ordering. This conjecture is further supported by the observation of sharp LEED pattern (Fig$.$ 1(b)), which following the six-fold symmetry of the (111) surface forms hexagonal motifs. In addition to intense principal spots, relatively weaker non-integer peaks are also discernable in the LEED image, which suggests presence of surface reconstruction. AFM characterization further reveals that the film surface is atomically flat (Fig$.$ 1(c)) and consists of well-defined step-terrace structure with a nominal step height of 0.23 nm that amounts to the one-unit cell of SRO along the [111] direction. Transport measurement shows a metallic behavior (Fig$.$ 1(d)) down to 2 K, along with a “kink” at 127 K that is characteristic of the onset of ferromagnetic phase transition. The residual resistivity $\sim$200 $\mu\Omega$ cm (at 2 K) compares well to the values reported for SRO (001) films of similar thickness  \cite{SRO_thickness}. Overall, the structural and electrical characterization demonstrates that high-quality SRO films can be grown on the (111) STO substrate.

Next, to probe the electronic band structure of  (111) SRO  films, we measured APRES on SRO films with thicknesses varying from 7 u$.$c$.$, 15 u$.$c$.$, 30 u$.$c$.$, and 50 u$.$c$.$.  While the thinnest film turns out to be insulating, ARPES measurements on the other three samples reveal a metallic nature with a sharp Fermi cutoff (Fig$.$ S2(a) in Supplemental Material \cite{supp}). Furthermore, the 15 u$.$c$.$ thick SRO film exhibits relatively sharper bands and clear Fermi surface than the thicker samples. In the main text, we, therefore, limit our discussion to the ARPES measurement performed on the 15 u$.$c$.$ thick SRO film.

\begin{figure}[t]
	\centering
	\includegraphics[scale=0.8]{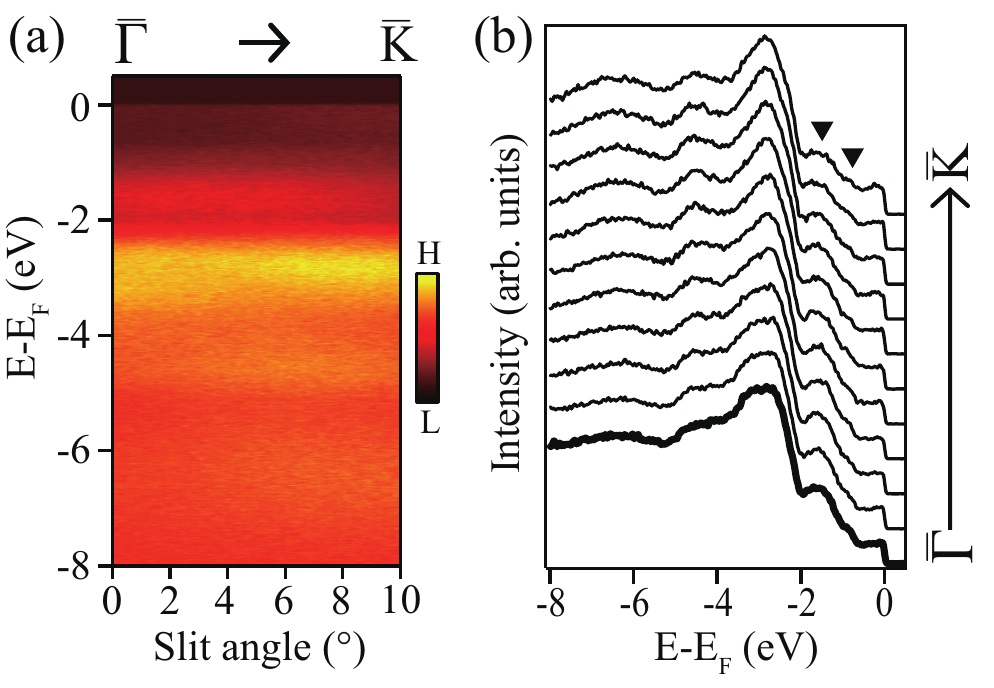}
	\caption{(Color online) (a) ARPES data over a wide energy window along the $\bar{\varGamma}$$-$$\bar{K}$ direction. (b) Corresponding angle-dependent EDC curves. Two triangles indicate the -1.4 and -0.8 eV peaks.}
	\label{fig:digraph}
\end{figure}
Figure 2(a) and 2(b) show the ARPES intensity plot along the high symmetry $\bar{\varGamma}$$-$$\bar{K}$ direction of the Brillouin zone (BZ) (Fig$.$ 3(a)) and the corresponding angle-dependent energy distribution curves (EDC), respectively. The valance band spectra show weak dispersion along this high symmetry direction. Nonetheless, the characteristic features associated with the O 2$p$ nonbonding and bonding states between -3 eV and -7 eV \cite{Shen, SRO_Japan} are discernable (Fig$.$ S3 in Supplemental Material \cite{supp}). Meanwhile, between -2 eV and the Fermi level, EDC displays two unconventional peaks centered around -1.4 and -0.8 eV (marked by the triangles), alongside a considerably suppressed quasiparticle (QP) peak at the Fermi level. We found these features are common to all (111) SRO films, irrespective of their thicknesses (Fig$.$ S2 in Supplemental Material \cite{supp}).
\begin{figure*}[t]
	\centering
	\includegraphics[scale=0.75]{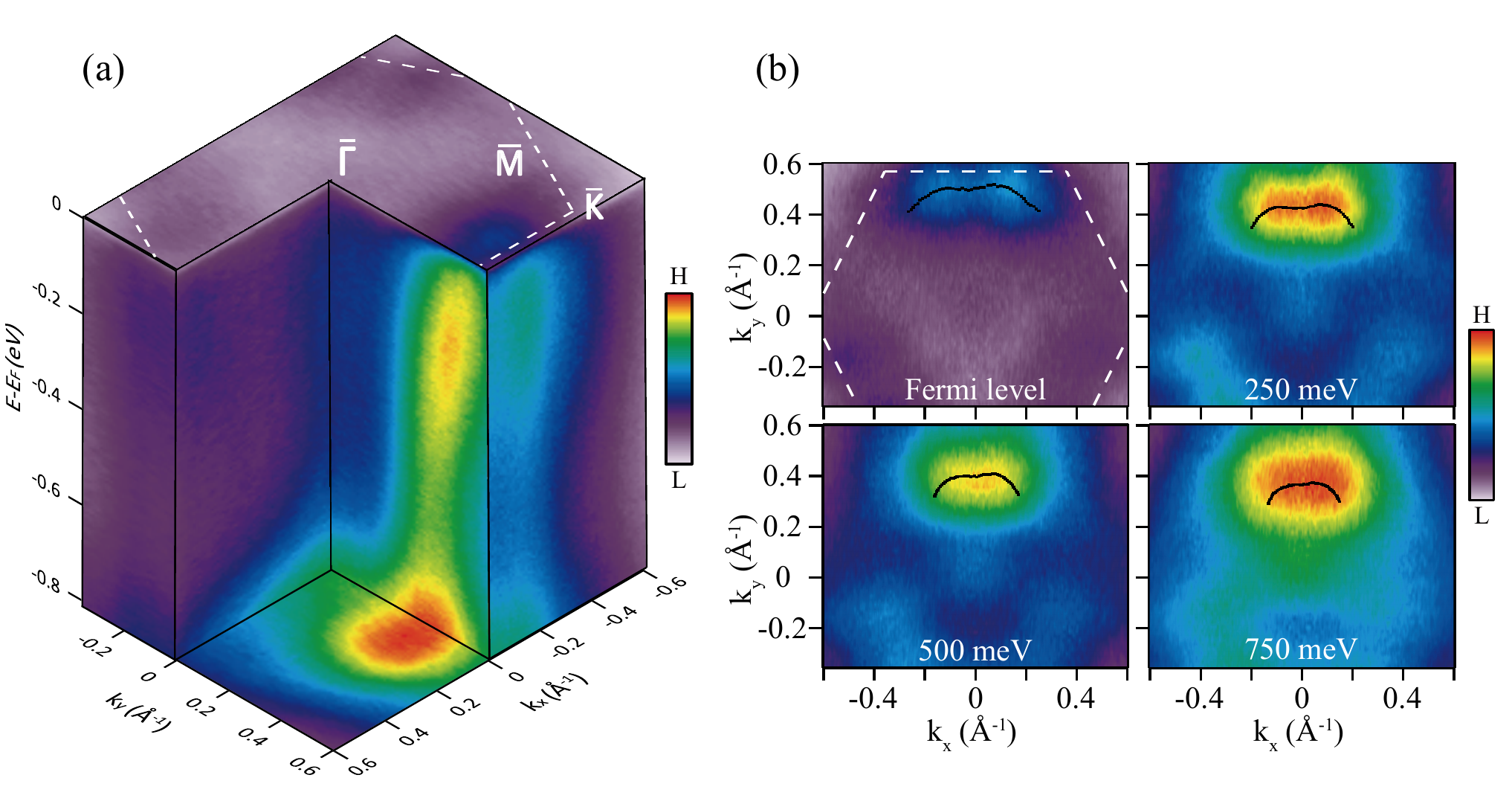}
	\caption{(Color online) (a) Three-dimensional energy versus momentum dispersion of the 15 u$.$c$.$ SRO film. The  dashed white line indicates the surface projected BZ. (b) Iso-energy surfaces at energies 0 meV (Fermi level), -250 meV, -500 meV and -750 meV, respectively. These iso-energy  surfaces  are obtained by integrating over an energy  window of $+10$ to $-10$ meV around the corresponding energies. The black dots are guides for the eyes, marking the coordinates of highest hotspot intensity that are obtained by profiling the iso-energy  surfaces within the azimuthal angular range of $+30^{\circ}$ to $-30^{\circ}$.}
	\label{fig:digraph}
\end{figure*}

To comprehend the origin of the suppressed QP intensity and the -1.4 eV peak, we compared the spectral weight of this peak for films with thickness ranging from 15-50 u$.$c$.$ and found it to be thickness-independent (Fig. S2. in Supplemental Material \cite{supp}). The residual resistance ratio (RRR), obtained from the transport measurements, however, increases with increasing thickness. Assuming that the RRR value inversely correlates to the Ru vacancy concentration in the SRO film, we, therefore argue that excessive Ru deficiency \cite{Shen,Off_cation} can not account for the -1.4 eV peak. Additionally, we studied an SRO film that was identically grown on the (001) STO substrate. The RRR value of this film ($\sim$4) is slightly larger than the 15 u.c. thick (111) film ($\sim$2.5), but the Curie temperature ($\sim$127 K) is identical. The valance band spectrum of this (001) SRO film exhibits a sharp QP peak, and void of any additional peaks down to 2 eV from the Fermi level (Fig$.$ S3 in Supplemental Material \cite{supp}), which further support our conjecture. Next, we consider disorders and enhanced electronic correlation, which can transfer spectral weight from the Fermi level to the so-called in-gap states and lower Hubbard band, respectively$-$yielding an incoherent peak around -1.3 eV \cite{SRO_CRO_ARPES, SJOh}. It is reasonable to expect that structural or compositional disorders could be present on the polar (111) surface as a means of compensating its polar charge. The single peak structure of the O $1s$ XPS spectrum (Fig$.$ S1(a) in Supplemental Material \cite{supp}), however, suggests that the contribution from the compositional disorders is minimal \cite{SJOh}. Structural disorders, namely, atomic reconstruction (evident in the LEED image Fig$.$ 1 (b)) and relaxation, therefore, naturally appear as the plausible driving mechanism. Understanding whether the structural disorders induce in-gap states or the lower Hubbard band requires further study, which is beyond the scope of this work. 

Having examined the valance band of the (111) SRO film, we turn our attention to the fermiology.  Figures 3(a) and (b) display the 3D ARPES image and iso-energy surfaces covering binding energies down to -0.8 eV from the Fermi level. The Fermi surface consists of three hotspot-pairs centered around the $\bar{M}$ points, forming six waterfall-like bands. The intensity of hotspots (or bands) exhibits azimuthal-angle dependence
(Fig$.$ S5 in Supplemental Material \cite{supp}). This suggests the waterfall-like bands  have a strong orbital character \cite{Azi_1,Azi_2} that is possibly caused by the degeneracy-lifting of the Ru $t_{2g}$ states under the trigonal crystal field imposed by the (111) orientation \cite{Okamoto_111}. Furthermore, the Fermi surface exhibits a three-fold symmetry, which differs from the expected six-fold symmetry of the BZ (marked by the white dashed lines in Fig$.$ 3(a) and (b)) projected on the (111) surface. This implies that for the given photon energy of 21.2 eV, we probe the three dimensional bulk BZ away from the high-symmetry points ($\varGamma$ or $Z$) along the $k_{z}$ axis. Assuming a nominal inner potential value of about 14 eV \cite{CRO_Shen}, we estimate that we probe the bulk BZ around $k_{z}$ $\sim$ $0.15$ $\pi$/$c$, where $c = \sqrt{3}a_{o}$ with $a_{o}$ (= 3.93 \AA) being the pseudo-cubic lattice parameter of SRO.

With the increasing binding energy, the hotspot-pairs shift towards the center of the BZ ($\bar{\varGamma}$). To better elaborate this shift, in Fig$.$ 3(b) we mark the coordinates with the highest hotspot intensity by black dots. This binding energy-dependent shift of the hotspot-pairs reflects the dispersive nature of the waterfall-like bands, which is more prominent in the 3D ARPES image in Fig$.$ 3(a). Tracking the band dispersion in Fig$.$ 3(a) and (b), further reveals that the photoemission intensity nonmonotonically varies with the binding energy. First, it increases from the Fermi level to about -0.25 eV, followed by a dip through -0.5 eV before peaking at -0.8 eV. At binding energies higher than -0.8 eV, we could not observe a clear band dispersion due to overlap with the signal from the non-dispersive  -1.4 eV feature. From these observations, we conclude that the high photoemission intensity at -0.8 eV can be attributed to the bottom of the bands, which leads to the -0.8 eV peak in the valance band spectra (Fig$.$ 2(b)). In addition to the  waterfall-like bands, we also observed a relatively weaker feature at the $\bar{\varGamma}$ point that vertically disperses down to -0.8 eV (Fig$.$ S4 in Supplemental Material \cite{supp}). At present, however, we do not understand the origin and implications of this vertically dispersive feature.

Next, to estimate the effective band mass and to gain insight into the many-body interactions, we further analyzed the waterfall-like band near the Fermi level. In Fig$.$ 4(a) we show the band dispersion down to -0.4 eV, which is extracted along the cut marked by the white line in the inset figure. From the curvature plot (Fig$.$ 4(b)) \cite{2ndDerivative}, which magnifies the dispersion between 0.4 to 0.52 \AA$^{-1}$ (dashed rectangle in Fig$.$ 4(a)), we evaluated the corresponding Fermi wave vector to be about 0.51 \AA$^{-1}$. Meanwhile, from the Lorentzian fitting of the momentum distribution curves (MDC) we obtained the MDC peak positions. The MDC peak dispersion is plot using red circles in Fig$.$ 4(c). With these information in hand, a quadratic polynomial fit to the high-binding energy part of the MDC peak dispersion (shown by the solid blue line in Fig$.$ 4(c)), yields a bare band mass of about $m_{b} = 0.41 \pm 0.02 m_e$. In contrast, by fitting the dispersion  within an energy range of $\pm$ 10 meV around the Fermi level, we obtained an effective quasiparticle $m^{*} = 0.76 \pm 0.04 m_e$. Therefore, the mass renormalization factor can be estimated to be about $m^{*}/m_b=1.85$ $\pm$ 0.13. Alternatively, from the ratio of the bare band velocity ($v_{b}$) to the quasiparticle velocity ($v^{*}$) we estimated the renormalization factor to be about of about 1.59 $\pm $ 0.13 (Fig$.$ S6 in Supplemental Material\cite{supp}). The comparable $m^{*}/m_b$ and $v_b/v^{*}$ values indicate the consistency of our approach. Interestingly, the renormalized quasiparticle mass ($m^{*}$) in (111) SRO is much lower than the values previously reported for (001) SRO films or layered Sr$_2$RuO$_4$, which nominally lie in the range 4-16$m_{e}$, and known to be strongly band dependent \cite{Shen, Tamai}. Recently, both light and heavy bands with $m_{b}$ values of 1$m_{e}$ and 14$m_{e}$, respectively, have been found to coexist in CaRuO$_3$ \cite{CRO_Shen}; these numbers are still larger than the band mass we obtained in this study. Although the extremely light band observed in (111) SRO films is surprising, it could be a natural consequence of probing a specific part of BZ. This limitation perhaps also hinders observing other heavier bands.

\begin{figure*}[ht!]
	\centering
	\includegraphics[scale=.85]{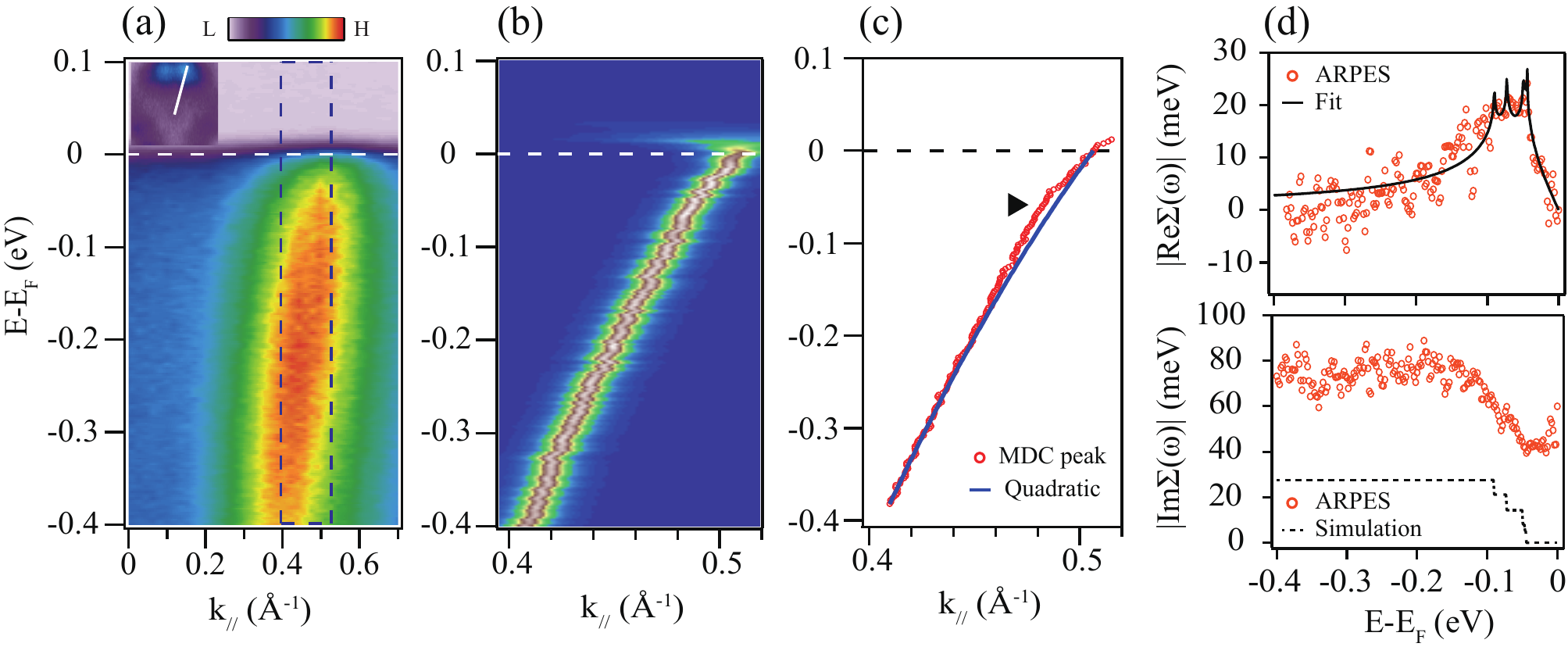}
	\caption{(Color online) (a) Energy versus momentum dispersion along the  solid white line in the inset figure. (b) 2D curvature data zooming the marked rectangular region in (a). (c) The MDC dispersion (Red circles) obtained by the Lorentzian fitting of MDC curves. The quadratic polynominal fit to the MDC is shown by the blue line. Black triangle indicates the ``kink" in the dispersion. (d) The orange circles plot the real part, Re$\Sigma(\omega)$ (upper panel) and the imaginary part Im$\Sigma(\omega)$ (lower panel) of the self-energy. The solid (dashed) black line mark the fit (simulation) of Re$\Sigma(\omega)$ (Im$\Sigma(\omega)$) using the Einstein model. The simulation is performed using parameters that are obtained by fitting Re$\Sigma(\omega)$.}
	\label{fig:digraph}
\end{figure*}

The observation of quasiparticle mass renormalization suggests that the electrons couple with Bosons, and the signature of this coupling can be found in the band dispersion. As evident in Fig$.$ 4(c), the MDC dispersion deviates from the quadratic behavior between about 100 meV to 44 meV (marked by a triangle). For SRO, the observation of this so-called ``kink" is attributed to the electron-phonon coupling \cite{Origin_of_kink}. To further support this assignment, in Fig$.$ 4(d) we show the real part (Re$\Sigma(\omega)$) and imaginary part (Im$\Sigma(\omega)$) of the self-energy that are calculated from the MDC. The real part of the self-energy (upper panel in Fig. 4(d)) exhibits a broad maximum covering an energy range similar to that of the ``kink''. To quantitatively analyze the underlying electron-phonon coupling, we considered the Einstein model, which accounts for the coupling between electrons and optical phonons \cite{Debye_Hufner, Debye_2005, Debye_2009, Einstein_Park, Einstein_Be, Debye_Tamai}. Assuming an effective electron-phonon coupling constant, $\lambda$ = 0.3 and five optical phonon modes with energies $\hbar$$w_{1}$ $=$ $44.04$ meV, $\hbar$$w_{2}$ $=$ $46.4$ meV, $\hbar$$w_{3}$ $=$ $49.5$ meV, $\hbar$$w_{4}$ $=$ $72.7$ meV, and $\hbar w_{5}$ $=$ $90.44$ meV\cite{Behera}, we can fit Re$\Sigma(\omega)$, as shown by the solid black line in the upper panel of Fig$.$ 4(d). The imaginary part of the self-energy (lower panel of Fig. 4(d)), which is linked to the scattering rate of electrons, gradually increases from 44 meV and attains a plateau above 100 meV. Unlike Re$\Sigma(\omega)$, the Einstein model, however, could not describe accurately the Im$\Sigma(\omega)$. Although the simulation reproduces the increase in Im$\Sigma(\omega)$ (dashed line in the lower panel of Fig. 4(d)), it could not account for a constant offset amounting about $\sim$45 meV. These discrepancies may be attributed to the additional contributions arising from the electron-defects scattering\cite{Debye_2009, Debye_e, Debye_Chulkov}. Nevertheless, the quantitative self-energy analysis demonstrates that the coupling between electrons and five optical phonon modes give rise to the ``kink" in the band dispersion.

The relation between the electron-phonon coupling constant $\lambda$ and the renormalization factor can be expressed as $m^{*}/m_b$ = $v_b/v^{*}$ = $(1+\lambda)$. Based on the polynomial fitting of MDC, therefore, $\lambda$ should be in the range of 0.6-0.85, which is fairly comparable to the value (=0.9) reported for the (001) SRO film \cite{Shen}. Simulating the self-energy according to the Einstein model, with $\lambda$ values of 0.6 and 0.85, however, leads a large difference between the data and calculation (please see Fig. S7 and associated discussion in Supplemental Material\cite{supp} for details). The discrepancy between $\lambda$ values obtained using Einstein modeling of the self-energy and polynomial fitting of MDC could arise due to the oversimplified assumption we made in the former approach; namely, the coupling strength to all phonon modes is identical. Therefore, we argue that $\lambda$ = 0.85 should be treated as the upper bound of the electron-phonon coupling constant and might be envisaged as an effect of condensing five phonon modes' contributions into a single one.

In summary, we have demonstrated that high quality SRO film can be grown along the polar (111) direction using the PLD technique. \textit{In situ} ARPES study reveals the existence of light bands in the (111) SRO film. The effective mass analysis yields a renormalized quasiparticle effective mass of $\sim$0.8$m_{e}$, which is lowest among the Ruthenates. The mass renormalization can be attributed to the coupling between electron and multiple phonon modes that yields characteristic ``kink'' in the band dispersion that spans an energy range between 100 to 44 meV. Also, we found that the quasiparticle spectral weight is suppressed at the Fermi level, and an incoherent peak appears at -1.4 eV, which we suggest possibly originating from the structural disorders that could be present on the polar (111) surface.  

This work also leaves some open questions and scope for future studies. For example, we could not identify the orbital character of the observed bands, nor could we clarify the origin of the vertical feature at the Brillouin zone center. Synchrotron-based ARPES measurements with variable polarization and photon energies, complemented by theoretical calculations, could allow comprehensively understanding the overall band structure, including the orbital character of the band and the vertical feature. Nevertheless, we hope that our work would further stimulate studies on (111) thin films of correlated oxides with strong spin-orbit coupling strength and eventually pave the way towards realizing novel topological quantum phases.

\begin{acknowledgments}
This work was supported by the Institute for Basic Science in Korea, Grant No.IBS-R009-G2 and IBS-R009-D1.
S.D. conceived and designed the research under the direction of T.W.N.. S.D. grew and characterized samples assisted by J.R.K., S.H., B.S., and D.K.. B.K., S.D., W.J.K., Y.K., S.I., and Z.Y. performed transport measurements. H.R., S.D. and B.K. studied the optimized sample cleaning condition. H.R. performed the ARPES and XPS measurements under the supervision of C.K. and with assistance from I.S., M.K., S.H., J.J., S.H. and W.K.. H.R. and Y.I. analyzed the data. H.R., S.D., Y.I., and C.K. discussed and interpreted the result. H.R. and S.D. wrote the manuscript with inputs from all the authors.
\end{acknowledgments}


\end{document}


\title{Supplemental Information:\\
	Electronic band structure of (111) SrRuO$_{3}$ thin film$-$an  angle-resolved photoemission spectroscopy study}

\author{Hanyoung Ryu}

\affiliation{Center for Correlated Electron Systems, Institute for Basic Science (IBS), Seoul 08826, Republic of Korea}
\affiliation{Department of Physics and Astronomy, Seoul National University (SNU), Seoul 08826, Republic of Korea}

\author{Yukiaki Ishida}
\affiliation{Center for Correlated Electron Systems, Institute for Basic Science (IBS), Seoul 08826, Republic of Korea}
\affiliation{Institute of Solid State Physics, The University of Tokyo, Kashiwa, Chiba 277-8581, Japan}

\author{Bongju Kim}
\author{Jeong Rae Kim}
\author{Woo Jin Kim}
\affiliation{Center for Correlated Electron Systems, Institute for Basic Science (IBS), Seoul 08826, Republic of Korea}
\affiliation{Department of Physics and Astronomy, Seoul National University (SNU), Seoul 08826, Republic of Korea}

\author{Yoshimitsu Kohama}
\affiliation{Institute of Solid State Physics, The University of Tokyo, Kashiwa, Chiba 277-8581, Japan}

\author{Shusaku Imajo}
\affiliation{Institute of Solid State Physics, The University of Tokyo, Kashiwa, Chiba 277-8581, Japan}

\author{Zhuo Yang}
\affiliation{Institute of Solid State Physics, The University of Tokyo, Kashiwa, Chiba 277-8581, Japan}

\author{Wonshik Kyung}

\author{Sungsoo Hahn}

\author{Byungmin Sohn}
\affiliation{Center for Correlated Electron Systems, Institute for Basic Science (IBS), Seoul 08826, Republic of Korea}
\affiliation{Department of Physics and Astronomy, Seoul National University (SNU), Seoul 08826, Republic of Korea}

\author{Inkyung Song}
\affiliation{Center for Correlated Electron Systems, Institute for Basic Science (IBS), Seoul 08826, Republic of Korea}

\author{Minsoo Kim}

\author{Soonsang Huh}

\author{Jongkeun Jung}

\author{Donghan Kim}
\affiliation{Center for Correlated Electron Systems, Institute for Basic Science (IBS), Seoul 08826, Republic of Korea}
\affiliation{Department of Physics and Astronomy, Seoul National University (SNU), Seoul 08826, Republic of Korea}

\author{Tae Won Noh}
\email{twnoh@snu.ac.kr}
\affiliation{Center for Correlated Electron Systems, Institute for Basic Science (IBS), Seoul 08826, Republic of Korea}
\affiliation{Department of Physics and Astronomy, Seoul National University (SNU), Seoul 08826, Republic of Korea}

\author{Saikat Das}
\email {DAS.Saikat@nims.go.jp} 
\thanks{Present address: Research Center for Magnetic and Spintronic Materials, National Institute for Materials Science, 1-2-1 Sengen, Tsukuba 305-0047, Japan}
\affiliation{Center for Correlated Electron Systems, Institute for Basic Science (IBS), Seoul 08826, Republic of Korea}
\affiliation{Department of Physics and Astronomy, Seoul National University (SNU), Seoul 08826, Republic of Korea}

\author{Changyoung Kim}
\email{changyoung@snu.ac.kr}
\affiliation{Center for Correlated Electron Systems, Institute for Basic Science (IBS), Seoul 08826, Republic of Korea}
\affiliation{Department of Physics and Astronomy, Seoul National University (SNU), Seoul 08826, Republic of Korea}

\maketitle
\newpage


\begin{figure}[!htb]
	\centering
	\includegraphics[scale=0.7]{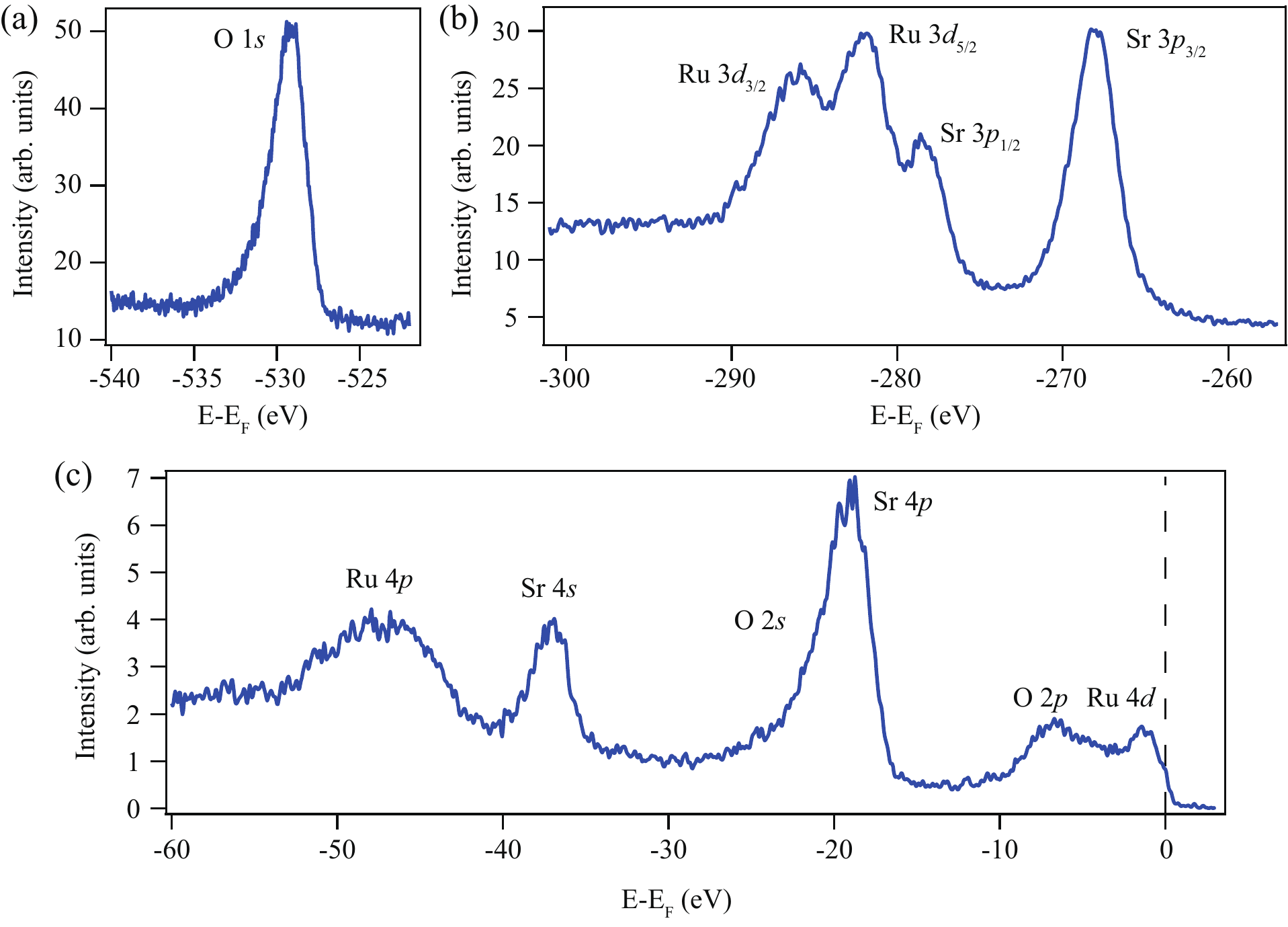}
	\captionsetup[figure]{labelformat=empty}
    \caption{(Color online) (a) XPS spectrum of O 1$s$. (b) XPS spectrum around Ru 3$d$ and Sr 3$p$ peaks. (c) XPS spectrum showing the Ru 4$p$, Sr 4$s$, Sr 4$p$, O 2$p$ peaks and valence band near the Fermi level. The Fermi level, which is calibrated using Au, is marked with a black dashed line.}

\end{figure}

Figure S1 shows the core-level spectra of the 15 unit cells (u$.$c$.$) thick (111) SRO film measured at room temperature. In Fig$.$ S1(a), O $1s$  spectrum shows a clear and sharp single peak centered around 529 eV, and the overall line shape is slightly asymmetric due to metallic screening \cite{Cardona}. Absence of shoulder peaks on the high binding side of the O $1s$ peak indicates the film surface is free from contamination such as hydroxide, carbonate, water, and also the amount of compositional disorders is minimal \cite{Ishida-san_STO, SRO_XPS, SJOh}. The core-level spectrum of Ru $3d$ and Sr $3p$ are shown in Fig$.$ S1(b). In Fig$.$ S1(c), we show the XPS spectrum covering the Ru $4p$, Sr $4s$, Sr $4p$, O $2p$, and Ru $4d$ peaks. The peak positions are consistent with the previous study \cite{SRO_XPS}. Please note that the energy resolution of the analyzer is $\sim$1 eV; this results in an enhanced broadening of the core level peaks. 

\newpage
\begin{figure}[!htb]
	\centering
	\includegraphics[scale=0.75]{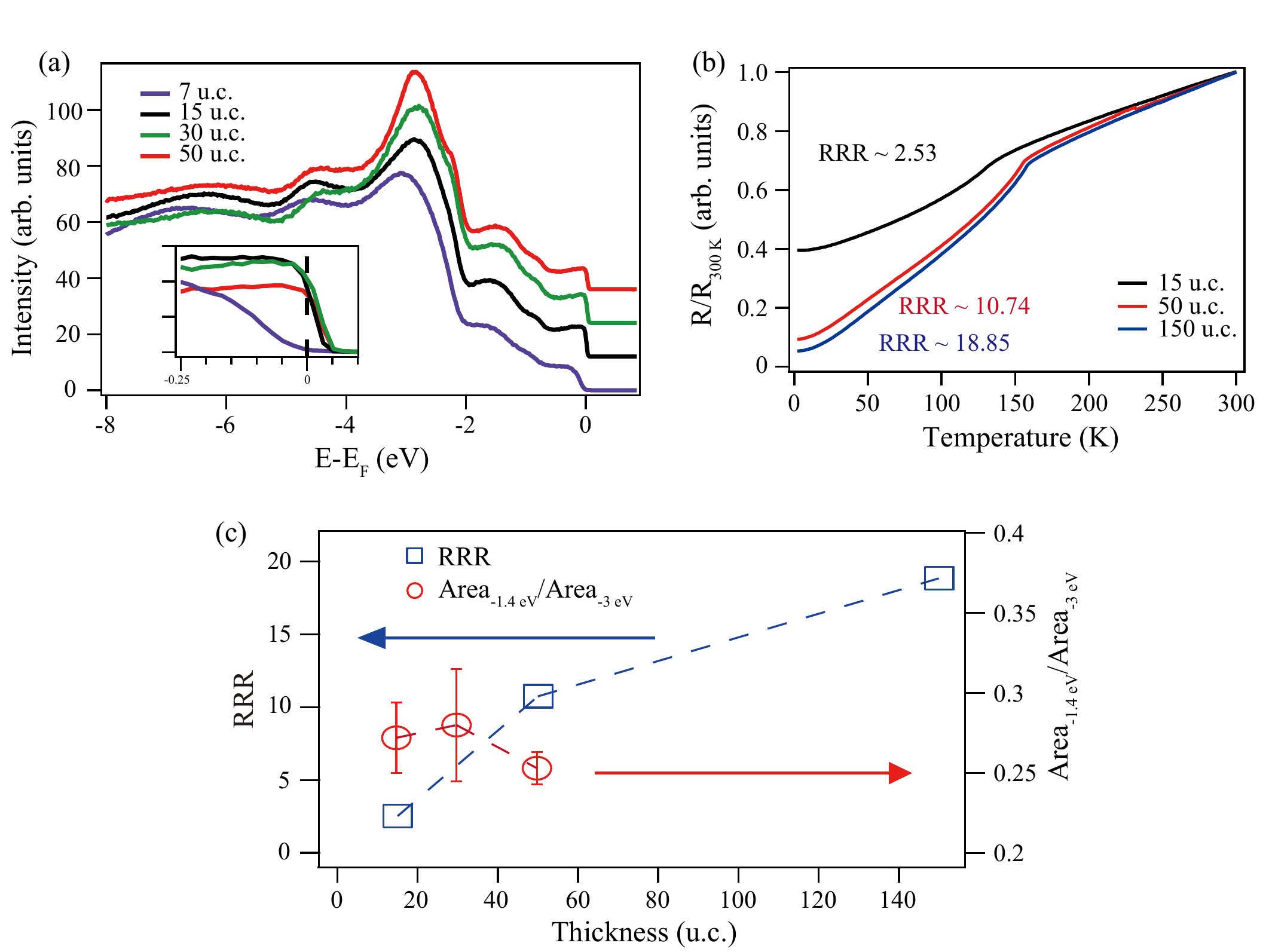}
	\caption{(Color online) (a) Angle integrated valence band spectra from 7, 15, 30, and 50 u$.$c$.$ (111) SRO films. The inset figure magnifies the spectra between the Fermi level and -0.2eV. The Fermi level that is calibrated using Au is marked with a black dashed line. (b) Temperature dependent resistance of 15, 50, 100 u$.$c$.$ (111) SRO films. The data are normalized to the room temperature (300 K) resistance value. (c) Residual resistance ratio and the spectral weight of -1.4 peak, which is normalized to the spectral weight of -3 eV as a function of film thickness.}
	\label{fig:digraph}
\end{figure}

Figure S2(a) shows the valence band spectra of the 7, 15, 30, and 50 u$.$c$.$ thick (111) SRO films taken at 20 K with intensity offset for clearance. For a clear comparison, the intensity of all four spectra is normalized with respect to the O $2p$ nonbonding peak intensity at binding energy 3 eV. Spectra from 15, 30, and 50 u$.$c$.$ sample show sharp Fermi cutoff at the Fermi level (inset of Fig$.$ S2(a)), which is calibrated with reference to the poly Au spectrum. However, the spectrum of the 7 u$.$c$.$ thick sample, plotted in purple color, is blue-shifted to the higher binding region and does not exhibit a sharp Fermi cutoff. The inset of Fig$.$ S2(a), which magnifies the spectra near the Fermi level, better describe this blue shift and the absence of the Fermi cutoff. These observations imply an insulating nature of the 7 u$.$c$.$ (111) SRO film, $\sim$1.6 nm. This insulating nature is consistent with recently reported thickness-dependent transport study, which shows a metal-insulator transition in (111) SRO film below 2.7 nm \cite{MIT_111}. Notably, all films exhibit the -1.4 eV and -0.8 eV peaks.

Figure S2(b) shows the temperature dependent resistance data of 15, 50, and 150 u$.$c$.$ thick (111) SRO films measured between 2K and 300K. For comparison, the data are normalized to the room temperature resistance. With increasing film thickness, the residual resistance ratio (RRR) and the Curie temperature improve, reaching about 19 and about 155K, respectively in the thickest film. This trend is consistent with the previous reports on (001) SRO films \cite{Main_32, thickness_001}, and implies that the Ru vacancy concentration decreases with increasing film thickness.

Figure S2(c) compares the thickness dependence of the RRR and the spectral weight of the -1.4 peak in the valance spectra (Fig$.$ S2(a)) that is normalized to the spectral weight of the non-bonding (-3 eV) peak. As discussed above, the RRR values, which are shown in blue squares, increase with increasing film thickness. The normalized spectral weights of the -1.4 eV (red squares), however, barely vary with thickness. This observation, therefore, suggests that the -1.4 peak is an intrinsic feature of the (111) SRO films that can not be associated to the Ru deficiency in the film.

\newpage
\begin{figure}[h]
	\centering
	\includegraphics[scale=0.65]{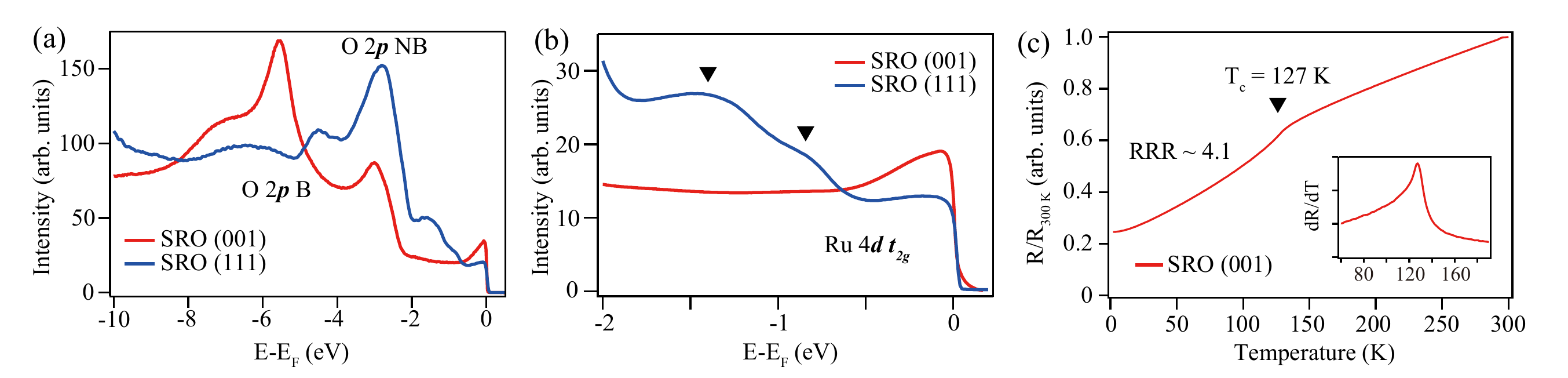}
	\caption{(Color online) (a) Angle integrated valence band spectra (001) and (111) SRO films, covering a binding energy up to 10 eV. Ru $4d$-O $2p$ bonding and O $2p$ non-bonding states are denoted by ‘B’ and ‘NB’, respectively. (b) Angle integrated valence band spectra magnifying the Ru $4d$ \textit{$t_{2g}$} dominated part of the valance band spectra within an energy window of 2 eV from the Fermi level. Please note that these spectra are extracted by summing over the whole momentum space, accessible in our experimental geometry. Black triangles mark the two peaks centered at -1.4 and -0.8 eV in the angle integrated spectrum of the (111) SRO film. (c) The temperature dependent resistance of the (001) SRO film, which is normalized to the room temperature resistance value. The black triangle mark the Curie temperature. The inset figure plots the derivative of the resistance with respect to temperature, highlighting the presence of ``kink'' coinciding with the ferromagnetic phase transition.}
	\label{fig:digraph}
\end{figure}
In Fig$.$ S3(a), we represent the angle integrated valence band spectra of 10 u.c. (~ 3.9 nm) thick (001) and 15 u.c. (~3.4 nm) thcik (111) SRO films. We integrated the energy versus slit angle spectrum along the slit angle direction. During the measurement, the slit sets to be parallel with $\bar{\varGamma}$$\textendash$$\bar{X}$ direction for (001) film and $\bar{\varGamma}$$\textendash$$\bar{K}$ direction for (111) film \cite{Shen}. Our (001) SRO spectrum well matches with reported results \cite{Shen, SRO_Japan}, and there are some different aspects between (001) and (111) results. Please see the main text for the detailed description.
In Fig$.$ S3(b), we plot the valance band spectra (within 2 eV from the Fermi level) that are summed over the momentum space covering $k_x$, $k_y$ = -0.4$\sim$0.4 \AA$^{-1}$. One can notice the two differences between (001) and (111) spectra. First, the intensity of the quasiparticle peak at the Fermi level of the (111) film is suppressed as compared to the (001) film. Second, while the (111) film exhibits clear peaks centered at -1.4 and -0.8 eV, the (001) SRO film does not show these peaks.
The resistance of the (001) SRO film, which is normalize to the room temperature resistance value, is presented in Fig. S3(c). The RRR value (~4.1) and the Curie temperature (~ 127K)) of this film is comparable to that of the 15 u.c. thick SRO (111) film. This implies that the nominal Ru content of these two films are comparable, which further support our argument that the -1.4 eV peak in the (111) film can not be linked to excessive Ru deficiency.

\newpage

\begin{figure}[h]
	\centering
	\includegraphics[scale=0.55]{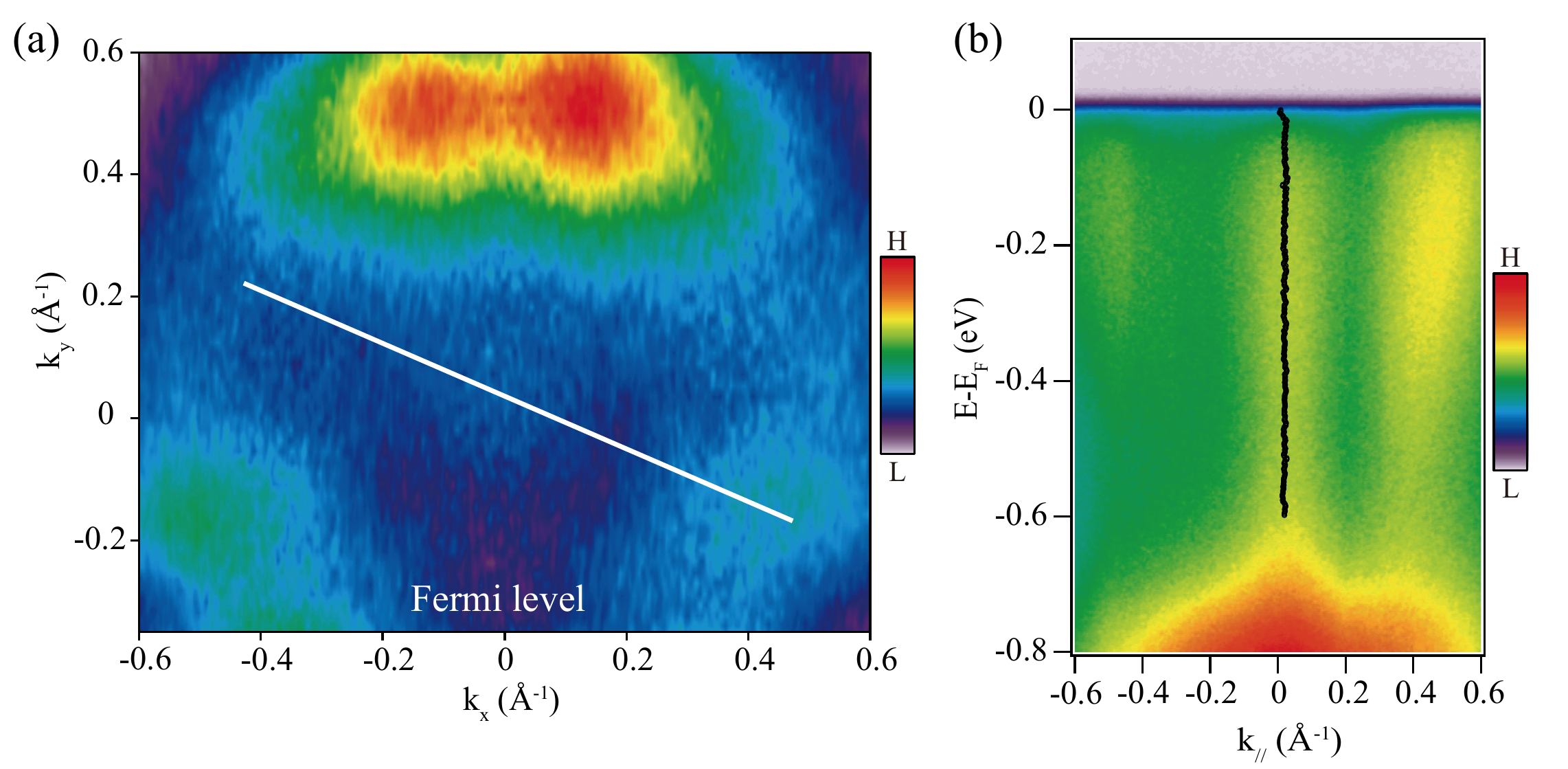}
	\caption{(Color online) (a) Fermi surface of the 15 u$.$c$.$ thick (111) SRO film, from Fig. 3(b) in the main text.
		(b) The energy versus momentum dispersion along the white solid line in Fig$.$ S4(a), displaying the vertical-dispersive feature alongside the water-fall like band.}
	\label{fig:digraph}
\end{figure}

Figure S4(a) shows the Fermi surface from the Fig$.$ 4(b) in the main text, but plot using different color contrast to make the dispersion at $\bar{\varGamma}$ more prominent. Fig$.$ S4(b) is the energy versus momentum dispersion including the vertical dispersive feature discussed in the main text. The peak position which is extracted from MDC of the band is presented with black dots in Fig$.$ S4(b). The band at $\bar{\varGamma}$ shows negligible dispersion from Fermi level to about -0.7 eV.

\newpage

\begin{figure}[h]
	\centering
	\includegraphics[scale=0.7]{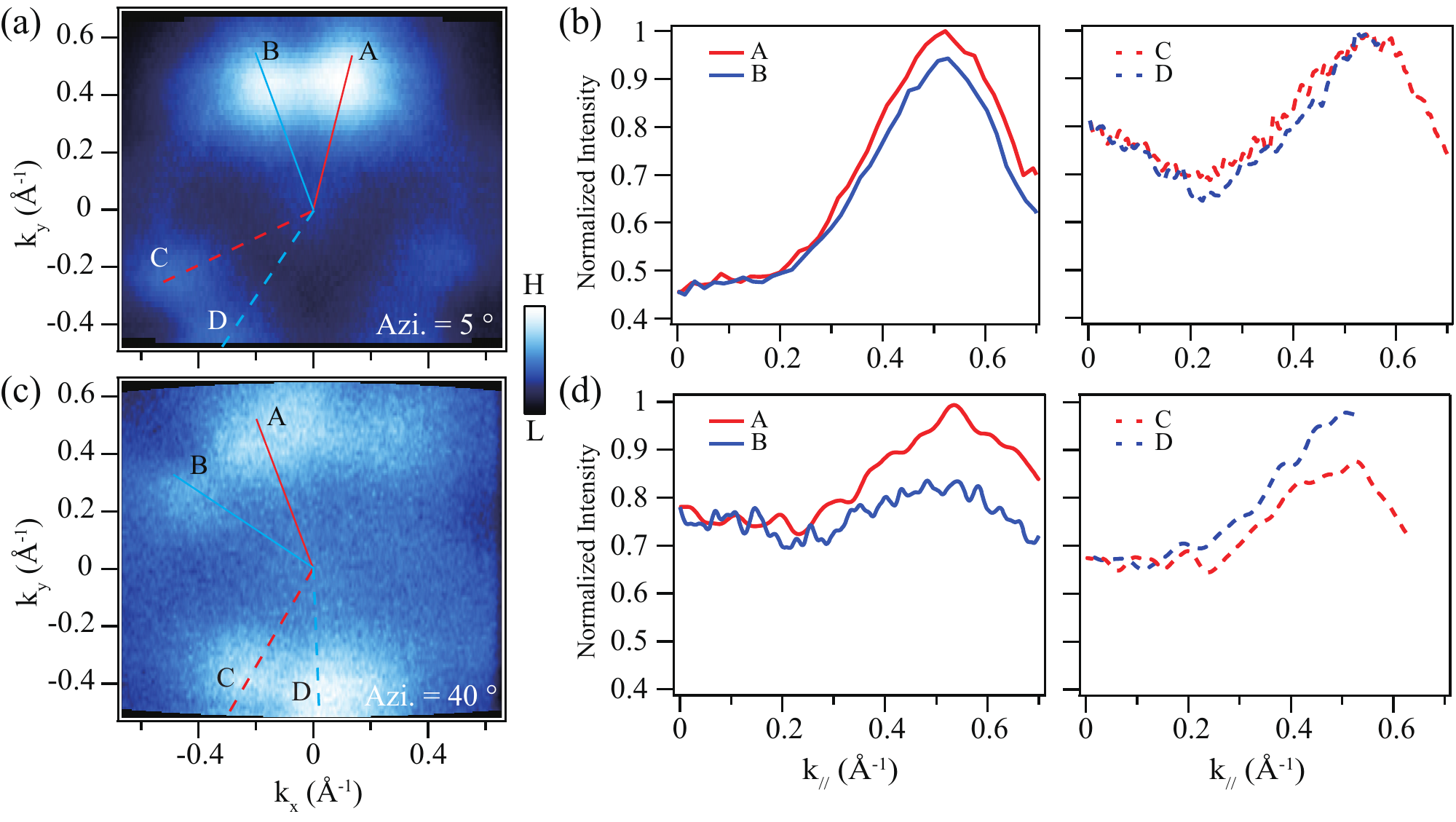}
	\caption{(Color online) (a) Fermi surface measured after rotating the sample azimuthally by 5 $^{\circ}$ counterclockwise. (b) (Left) Intensity versus momentum plot along the lines  A and B in (a). (Right) Intensity versus momentum plot along the lines C and D in (a). (c) Fermi surface measured after rotating the sample azimuthally by 40 $^{\circ}$ counterclockwise. (d) (Left) Intensity versus momentum plot along the lines A and B in (c). (Right) Intensity versus momentum plot along the C and D in (c).}
	\label{fig:digraph}
\end{figure}

We performed azimuthal dependent Fermi surface mapping and the results are displayed in Fig$.$ S5. Figure S5(a) and (c) is the Fermi surface measured after rotating the sample azimuthally by 5 and 40 degrees counterclockwise, respectively. In Fig$.$ S5(b) and (d), we show the intensity versus in-plane momentum plot extracted from (a) and (c), respectively. For a clear comparison, the maximum intensity is set to 1 in (b) and (d). In the case of 5 $^{\circ}$, intensity along the A (C) and B (D) show very close value in all in-plane momentum. But, in the case of 40 $^{\circ}$ rotation, the intensity of B is reduced by 20 $\%$ compared to A. Also, C and D show a similar tendency with A and B. One can notice that the hotspot-intensity is inversely proportional to their distance from the $k_y$ axis. This strong azimuthal angle-dependent intensity anisotropy or the so-called matrix element effect might imply an orbital selective occupation of the Ru $4d$ \textit{t$_{2g}$} band \cite{Azi_1, Azi_2}.

\newpage

\begin{figure}[h]
	\centering
	\includegraphics[scale=0.8]{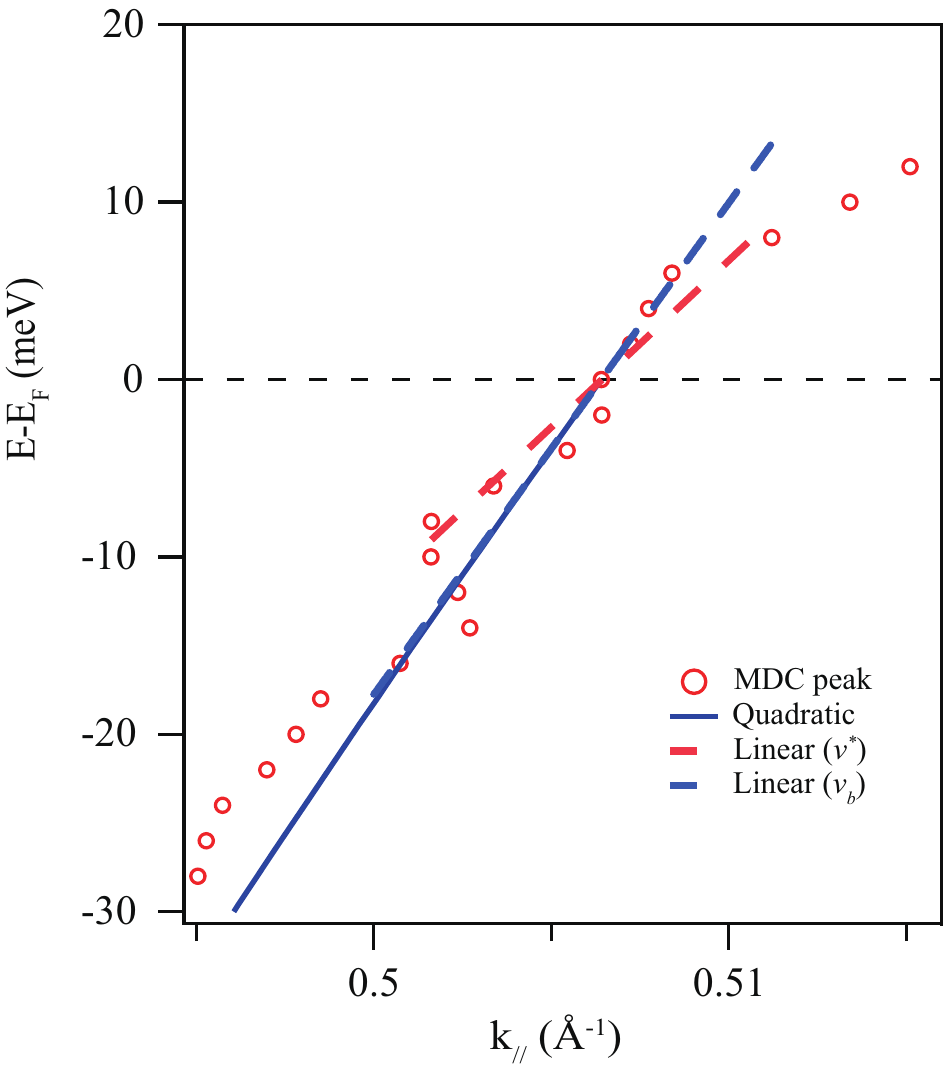}
	\caption{(Color online) The band dispersion near the Fermi level extracted from Lorentzian fitting of MDC curves is shown by red circles. The quadratic fitting, which we use to extract the bare band mass is shown by the solid blue line. The linear fitting in the energy range of -8 to 8 meV from our MDC peak (red circles) is shown by the dashed red line. The dashed blue line represents the linear contribution of the quadratic polynomial fitting, which we use to estimate the bare band velocity near the Fermi level ($v_{b}$).}
	\label{fig:digraph}
\end{figure}

Apart from extracting the bare band mass ($m_{b}$) and renormalized band mass ($m^{*}$), we also estimated the bare and renormalized velocities from the MDC dispersion. To calculate the renormalized band velocity at Fermi level, we performed the fitting with linear function at $E_{F}$ $\pm$ 8 meV from MDC peak (red circles). The calculated renormalized band velocity at Fermi level ($v^{*}$) is 1.76 $\pm$ 0.15 eV \AA. The calculated bare band velocity at Fermi level ($v_{b}$) obtained from the fitting is 2.77 eV \AA. From these results, we estimate $v_{b}/v^{*}$ = 1.59 $\pm$ 0.13, which compares fairly well to the renormalized factor, $m^{*}/m_{b}=1.85$ $\pm$ 0.13.
\newpage

\begin{figure}[h]
	\centering
	\includegraphics[scale=0.8]{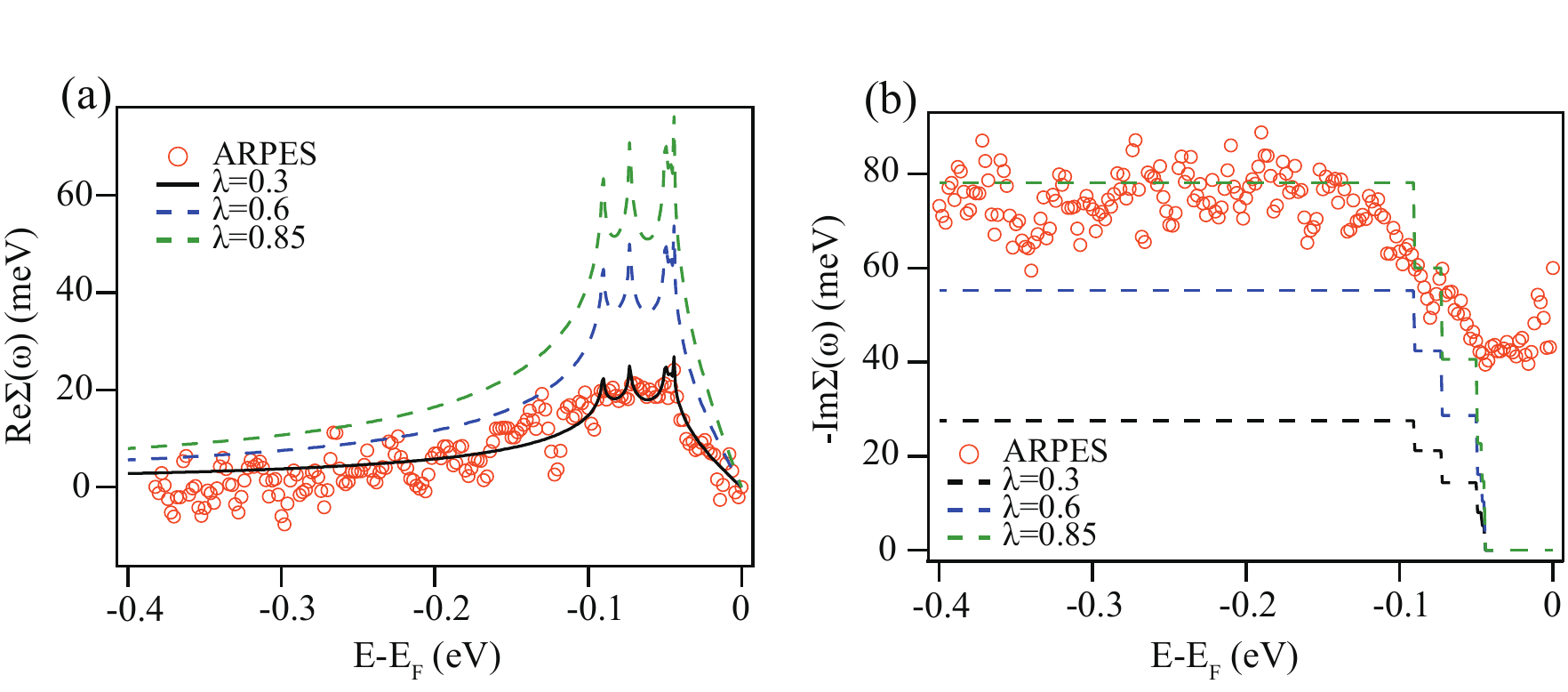}
	\caption{(Color online) (a) and (b) show the real part (Re$\Sigma$($\omega$)) and the imaginary part (Im$\Sigma$($\omega$)) (orange circles) of the self-energy that are calculated from the MDC dispersion in Fig. 4(c) in the main text. The fitting and simulation according to the Einstein model are shown in solid and dashed lines, respectively.}
	\label{fig:digraph}
\end{figure}

To quantitatively describe the electron-phonon coupling, we performed Einstein modeling of the self-energy. The rational behind this choice  is derived from the consideration that the ``kink" in the dispersion originates from the coupling of electrons to optical phonons \cite{Einstein_Park}. Whereas the Debye model accounts for coupling to the acoustic phonons\cite{Debye_2005, Debye_Tamai}, the Einstein model is suitable for the optical phonon modes \cite{Einstein_Park, Einstein_Be}. We considered five optical phonon modes with energies $\hbar$$w_{1}$ $=$ $44.04$ meV, $\hbar$$w_{2}$ $=$ 46.4 meV, $\hbar$$w_{3}$ $=$ 49.5 meV, $\hbar$$w_{4}$ $=$ 72.7 meV, and $\hbar$$w_{5}$ $=$ 90.44 meV \cite{Behera}. In the model, the Eliashberg function accounting for the electron-phonon coupling takes the form, $\alpha^{2} F(\omega) = 1/2 \lambda \omega_{E}\delta (\omega - \omega_{E})$ \cite{Debye_2005}. Here, $\lambda$ and $\omega_{E}$ are the electron-phonon coupling constant and characteristic phonon mode frequency, respectively.
 
For simplicity, we assumed that the coupling strength to all phonon modes is identical. With this assumption, a least-square fitting is perfromed to the real part of the self-energy (Black solid line in Fig. S7(a)), yielding  $\lambda$ = 0.3, and the contribution of phonons with energies $\hbar$$w_{1}$ $=$ $44.04$ meV, $\hbar$$w_{2}$ $=$ 46.4 meV, $\hbar$$w_{3}$ $=$ 49.5 meV, $\hbar$$w_{4}$ $=$ 72.7 meV, and $\hbar$$w_{5}$ $=$ 90.44 meV of about 0.25, 0.13, 0.27, 0.2, and 0.15, respectively. In addition, we also show the simulations with different $\lambda$ values ~ 0.6 (Blue dashed line) and 0.85 (Green dashed line), but fixing contributions of the phonons to the values estimated from the fitting, mentioned above. Concerning the imaginary part of the self-energy, we performed simulations with the $\lambda$ values 0.3 (Black dashed line), 0.6 (Blue dashed line), and 0.85 (Green dashed line), and the contribution from the phonon modes were set to the best fit values mentioned above (Fig.  S7(b)).